\documentstyle[twocolumn,prb,aps,epsf]{revtex}
\begin{document}
\draft

\twocolumn[\hsize\textwidth\columnwidth\hsize\csname @twocolumnfalse\endcsname

\title{
de Haas-van Alphen Effect in the Two-Dimensional and
the Quasi-Two-Dimensional Systems
}

\author
{Keita Kishigi}
\address
{Japan Science and Technology Corporation, Domestic Research Fellow, Japan \\
and Condensed Matter Physics Group, Nanoelectronics Research
Institute, AIST, Tsukuba, Ibaraki 305-8568, Japan}
\author
{Yasumasa Hasegawa}
\address
{Faculty of Science, Himeji Institute of Technology,
   Ako, Hyogo 678-1297, Japan}

\date{\today}
\maketitle

\begin{abstract}
We study the de Haas-van Alphen (dHvA)
oscillation
in two-dimensional and quasi-two-dimensional systems.
We give a general formula of the dHvA
oscillation in  two-dimensional multi-band systems.
By using this formula, the dHvA
oscillation and its temperature-dependence for the two-band system are shown.
By introducing the interlayer hopping $t_z$, we examine the crossover
from the two-dimension, where the oscillation of
the chemical potential plays an important role in the magnetization
oscillation, to the three-dimension, where the oscillation of the
chemical potential can be neglected as is well know as the  Lifshitz
and Kosevich formula.
The crossover is seen at $4 t_z \sim 8 ta b H /\phi_0$,
where $a$ and $b$ are
lattice constants, $\phi_0$ is the flux quantum and
$8t$ is the width of the total energy band.
We also study the dHvA oscillation in  quasi-two-dimensional
magnetic breakdown systems. The quantum interference oscillations such
as $\beta$-$\alpha$ oscillation as well as the fundamental
oscillations are  suppressed by the interlayer hopping $t_z$, while the
$\beta$+$\alpha$ oscillation gradually increases as $t_z$ increases
and it has a maximum at $t_z/t\approx 0.025$. This interesting
dependence on the dimensionality can be observed in the
quasi-two-dimensional organic conductors with
uniaxial pressure.
\end{abstract}

]


\narrowtext

\section{Introduction}
It is well known that
the magnetization ($M$) periodically oscillates as a
function of the inverse magnetic field ($1/H$) with frequencies
of the
extremes of the cross-section area of the Fermi surface.
This oscillation, known as the de Haas-van Alphen (dHvA)
oscillation, is used to study the shape of
the Fermi surface of the conductors 
and the experiments have been fitted by using the Lifshitz and Kosevich (LK)
formula\cite{LK,Shoenberg84}.
 From the fitting of the temperature-dependence
of the amplitude of the dHvA oscillation,
we can obtain the cyclotron effective mass ($m$)
for each cyclotron orbits.

Recently, a revival of interest in dHvA oscillations
in two-dimensional and quasi-two-dimensional conductors 
arises, since it has been noticed that
the LK formula can not be applied in the
two-dimensional
\cite{Shoenberg2d,machida,kishigi,harrison,kishigi2,nakano,alex,alex2,fortin,fortinprl}
and quasi-two-dimensional\cite{grigoriev} conductors.
Because of the Landau quantization of the two-dimensional  kinetic
energy in the magnetic field, both electron number ($N$) and  chemical
potential ($\mu$) cannot be fixed simultaneously as the magnetic field
is changed.
In the actual situation, $N$ is fixed, and $\mu$
oscillates periodically as a function
of $1/H$.
In the three-dimensional conductors the oscillation of $\mu$
   is shown to be very small\cite{LK,Shoenberg84}.
Therefore the LK formula obtained under the condition
of fixed $\mu$ (grand canonical ensemble
with treating $\mu$ independent
of $H$) is justified to be used in the system of
   the fixed  $N$
(canonical ensemble).
However, in the two-dimensional system,
the oscillation of $\mu$ as a function of $1/H$  cannot be neglected.
The saw-tooth waveform of the dHvA oscillation in the fixed $N$ case
is inverted from that in the fixed $\mu$ case
in the single-band case\cite{Shoenberg2d}.

It has been shown that there appear the quantum interference
oscillations with the frequencies such as
$f_{\beta}-f_{\alpha}$ and
$f_{\beta}+f_{\alpha}$ in the magnetization
when $N$ is fixed
  in  magnetic breakdown systems
of the tight-binding
electrons\cite{machida,kishigi,kishigi2,kim,han},
  in the magnetic-breakdown systems  treated
semi-classically\cite{harrison,fortin,fortinprl},
and in multi-band systems\cite{nakano,alex,alex2,nakano2}.
The anomalous
spin-splitting effects are also shown to exist in the fixed $N$
systems\cite{kishigispin,kishigispin2,nakanospin,itsspin}.
The  quantum interference oscillations
have been confirmed by the experiments of
the dHvA effect in various two-dimensional and quasi-two dimensional
systems such as
the quasi-two-dimensional organic conductors\cite{yamaji} ($\kappa$-(BEDT-TTF)$_2$Cu(NCS)$_2$\cite{Meyer,uji}
and  $\alpha$-(BEDT-TTF)$_2$KHg(SCN)$_4$\cite{honold}),
Sr$_2$RuO$_4$\cite{mck,ohmich,yoshida},
GaAs/AlAs heterointerface\cite{Deutschmann}
and In$_x$Ga$_{1-x}$As single
quantum well structure\cite{shep}.

 From numerical calculations for the
single-band systems\cite{harrison2,grig2}, it has been
shown that the temperature-dependence of the
amplitude of the dHvA oscillation
can not be described by the LK formula.
The conventional fitting of the LK
formula may not give  the cyclotron effective mass in the two-dimension.


In this paper we give a general formula for the
dHvA oscillation in
two-dimensional multi-band systems.
 From this formula,
the oscillatory parts of the chemical potential and the
magnetization are calculated numerically in the two-band system.
Moreover, by increasing the
interlayer hopping $t_z$,
we study how the dHvA oscillation in the two-dimensional system change to
those in the three-dimensional system.

In the next section we give the general formula for the dHvA oscillation.
In section \ref{sect:2d} we show the difference of dHvA oscillations in
cases of the fixed $\mu$ and the fixed $N$ in the two-dimensional
multi-band system.
In Section \ref{sect:q2d} the crossover from two-dimension to
three-dimension is studied.
In section
\ref{QIO} the quantum interference oscillation are studied in two-dimensional
and quasi-two-dimensional magnetic breakdown systems.

\section{Formulation}
The thermodynamic potential $\Omega$ is calculated
as a function of the chemical potential $\mu$, magnetic field $H$ and
temperature $T$ as
\begin{eqnarray}
\Omega(\mu,H)
&=&
-T \log \left(
\sum_{N,E} \exp \left(-\frac{E- \mu N}{T} \right)
   \right)
\nonumber \\
&=& -T \int d\epsilon N_0(\epsilon, H)
\log \left(1+\exp\left( -
\frac{\epsilon -\mu}{T} \right) \right),
\nonumber \\
\label{omega}
\end{eqnarray}
where
$E$ is the total energy of the system,
$N$ is the number of
electrons and
$N_0(\epsilon, H)$ is the density of states in the presence
of the magnetic field $H$.
The magnetization $M$ is obtained in the grand canonical ensemble as
\begin{equation}
   M_{\mu}(\mu, H) = -  \frac{\partial \Omega(\mu, H)}{\partial H} .
   \label{MmuH}
\end{equation}
As mentioned by Lifshitz and Kosevich\cite{LK}, $\mu$ is treated as
the independent variable of $H$ in the above equation. This is not the
case in the actual situation.

The electron number $N$ is calculated in the grand canonical ensemble
as
\begin{equation}
N(\mu, H)=-\frac{\partial \Omega(\mu, H)}
{\partial \mu}.
\label{functionN}
\end{equation}
If we take $\mu$ to be independent of $H$,
    $N$ oscillates as a function of $H$.
In order to study the system with constant $N$, i.e. canonical
ensemble,
we should consider Eq.(\ref{functionN}) as the equation
giving the chemical potential $\mu (N, H)$ as
a function of $N$, $H$ and $T$,
\begin{equation}
N= - \frac{\partial \Omega(\mu(N,H),H)}
{\partial \mu(N,H)}.
\label{eqNconst}
\end{equation}
In this case $\mu$ oscillates as a function of $H$.
Then we get the Helmholtz free energy
as
\begin{eqnarray}
F(N, H) &=& \Omega(\mu(N,H), H) + \mu(N,H) N
\label{helmholtz}
\end{eqnarray}
The magnetization should be calculated
in the canonical ensemble as
\begin{equation}
   M_{N}(N,H) = -   \frac{\partial F(N,H)}{\partial H} .
\label{eqmNh}
\end{equation}
Using Eq.(\ref{helmholtz}) we get
\begin{eqnarray}
M_{N}(N,H) &=& - \frac{d \Omega( \mu(N,H), H)}
{d H} - \frac{\partial \mu(N,H)}{\partial H}  N
\nonumber \\
    &=& -\left( \frac{\partial \Omega ( \mu, H)}{\partial H}
        \right)_{\mu=\mu(N,H)}
\nonumber \\
&=& M_{\mu}(\mu(N,H),H)
\label{MNH}
\end{eqnarray}
We see that the magnetization calculated in the
canonical ensemble has the same form as the
magnetization calculated in the grand canonical ensemble,
but we have to take into account the oscillation of $\mu$
in the system of the fixed electron number.
This is the general formula independent of
the dimension of the system.

\section{two-dimensional electrons}\label{sect:2d}
\subsection{multi-band system}
In this section we study the two-dimensional multi-band electrons in
the magnetic field.
This model can be applied to the multi-band materials such as
Sr$_2$RuO$_4$ and In$_x$Ga$_{1-x}$As.
The density of states in the magnetic field is given by
the sum of  delta functions,
\begin{equation}
N_0(\epsilon, H) = \sum_{j} \sum_{n=0}^{\infty}
    \sum_{\sigma=\pm 1/2}
\rho_j \omega_j \delta(\epsilon-\epsilon_{j n \sigma}),
\end{equation}
where $j$ is the band suffix, $\rho_j$ is the density of
states for each spin in the $j$ band at $H=0$,
\begin{equation}
\omega_j= \frac{e H}{m_j},
\end{equation}
is the
cyclotron frequency with the cyclotron mass $m_{j}$,
\begin{equation}
   \epsilon_{j n \sigma}=
   \omega_{j}(n+\frac{1}{2}) + g_{j}\sigma\mu_B H
+\Delta_{j},
\end{equation}
is the Landau level, $\Delta_{j}$ is the energy of the band bottom
   in zero magnetic field, $g_{j}$ is the electron
$g$-factor for the $j$ band
and  $\mu_B=e \hbar / 2 m_0 c_0$ is the Bohr magnetron with electron
mass $m_0$ and velocity of light
$c_0$.
In this section, we set
$\hbar=c_0=k_B=1$.
The electron number at $H=0$ is given by
\begin{eqnarray}
N&=&\sum_{j}\sum_{\sigma=\pm 1/2}
\int_{\Delta_{j}}^{\mu_0}\rho_{j} d\epsilon
=2\sum_{j} \rho_{j}(\mu_0- \Delta_{j}),
\label{totalN}
\end{eqnarray}
where $\mu_0$ is the chemical potential at $H=0$.

Applying the Poisson formula\cite{Shoenberg84} in
Eq.(\ref{omega}),
thermodynamic potential is written as\cite{alex3}
\begin{equation}
   \Omega(\mu,H)=
\Omega_0(\mu,H) + \tilde{\Omega}(\mu, H),
\label{omega0tilde}
\end{equation}
where
\begin{eqnarray}
   \Omega_0(\mu,H)&=&-T\int_0^{\infty}d\epsilon  \sum_{j}
   \sum_{\sigma=\pm
1/2} \rho_{j}
\nonumber \\
   &\times& \log \left( 1 + \exp\left(
\frac{\mu-\Delta_{j} -
g_{j}\sigma \mu_{B}H -\epsilon}{T}\right)  \right)
\label{Poisson}
\end{eqnarray}
and
\begin{eqnarray}
    \tilde{\Omega}(\mu, H)
   &=&\frac{1}{12} \sum_{j}
    \rho_{j} \omega_{j}^2
   \nonumber \\
   &+& \sum_{j}\sum_{r=1}^{\infty}
\frac{\rho_j\omega_j^2}{\pi^2 r ^2}\times
R_{T,j r} \times R_{S,j r}
   \nonumber \\
&\times&
\cos \left( 2\pi r \left( \frac{(\mu-\Delta_{j})}{\omega_j}
   + \frac{1}{2}
     \right)\right) ,
\label{eqtildeomega}
\end{eqnarray}
where
\begin{eqnarray}
R_{T,j r} &=&\frac{2 \pi^2 r T/\omega_{j}}
{\sinh (2 \pi^2 r T/\omega_{j})},
\label{RT}
\\
R_{S,j r} &=&\cos\left( r \pi \frac{\tilde{g}_j}{2} \right),
\label{RS}
\end{eqnarray}
and
\begin{equation}
   \tilde{g}_{j}= g_{j} \frac {m_{j}}{m_0}.
\end{equation}
In the above we have used the identity
\begin{eqnarray}
    \int_{0}^{\infty} \frac{\sin x}{1+e^{B x -A}} dx &=&
   1 - \frac{\pi \cos (A/B)}{B \sinh (\pi/B)}
    \nonumber \\
    & & -\sum_{n=1}^{\infty}\frac{(-1)^{n+1} e^{- A n}}{1+ B^2 n^2},
\end{eqnarray}
where $A=(\mu-\Delta_{j}-g_j\sigma \mu_B H)/T$,
   $B=\omega_j/(2 \pi r T)$, and we have neglected
the summation term over
$n$, since we are interested in low temperature properties ($A \gg 1$).

At low temperature ($T \ll \mu -\Delta_{j}$), we get
\begin{eqnarray}
   \Omega_0(\mu, H)
   &\approx& - \frac{1}{2} \sum_j \sum_{\sigma=\pm 1/2} \rho_j
(\mu - \Delta_{j} - g_j \sigma \mu_B H)^2 .
\label{omega0}
\end{eqnarray}
We obtain
\begin{equation}
   \frac{\partial \Omega_0 (\mu, H)}{\partial H}
    = - \frac{\mu_B^2}{4} H \sum_j \rho_jg_j^2.
\label{domega0/dh}
\end{equation}
 From this equation, it is found that
$\partial \Omega_0(\mu,H)/\partial H$ does
not make the oscillation of the magnetization as a
function of $1/H$.
Thus, in grand canonical ensemble,
$\tilde{\Omega}(\mu, H)$
   gives the oscillatory part of
the magnetization $\tilde{M}_{\mu}(\mu, H)$ as
\begin{eqnarray}
    \tilde{M}_{\mu}(\mu, H)&=& -\frac{\partial \tilde{\Omega}(\mu, H)}
    {\partial H} \nonumber \\
   &\approx& - \sum_{j}\sum_{r=1}^{\infty}
   \frac{2e^{2} f_j \rho_j }{\pi r m_j^{2}}
R_{T,j r}  R_{S,j r}
\nonumber \\
&\times&
\sin \left( 2\pi r \left( \frac{f_{j}}{H}
   + \frac{1}{2}
     \right)\right)
\label{eqmtildemu}
\end{eqnarray}
where
\begin{eqnarray}
2 \pi e f_{j}&=&2 \pi m_{j}(\mu-\Delta_{j}),
\end{eqnarray}
is  the area of the Fermi surface at $H=0$ in the $j$ band. Conventionally,
$R_{T,j r}$ is known as the temperature
reduction factor of the
LK formula and is used to estimate
the cyclotron effective mass ($m_j$) from the $T$-dependence
of the amplitude of the oscillation.
The electron number is obtained from Eq. (\ref{functionN}) as
\begin{eqnarray}
N(\mu,H) &=& -\frac{\partial \Omega_0}{\partial \mu} -\frac{\partial
\tilde{\Omega}}{\partial \mu}
    \nonumber \\
&=& \sum_{j} 2 \rho_{j} \biggl\{
   \mu - \Delta_{j}
   \nonumber \\
   & & +\sum_{r=1}^{\infty} \frac{\omega_j}{\pi r}
R_{T,j r} R_{S,j r}
\nonumber \\
&&\times\sin \left( 2\pi r \left(\frac{\mu-\Delta_{j}}{\omega_{j}}+
\frac{1}{2} \right) \right)
   \biggr\}.
\label{eqNmuh}
\end{eqnarray}

Here we consider the fixed $N$ case.
As we wrote in the previous section, the above equation
should be taken as the equation for $\mu(N,H)$.
With the use of Eq.(\ref{totalN}), we get,
\begin{eqnarray}
{\tilde \mu}(N,H)&=&\frac{-1}{\rho}\sum_{j}\sum_{r=1}^{\infty}
\frac{\rho_j \omega_j}{\pi r}R_{T,j r} R_{S,j r}
\nonumber \\
   & &\times
\sin \left( 2\pi r \left(\frac{{\tilde \mu}(N,H)
   +\mu_0-\Delta_{j}}{\omega_{j}}+
\frac{1}{2} \right) \right),
\label{mutilde}
\end{eqnarray}
where
\begin{eqnarray}
\rho=\sum_{j}{\rho}_{j},
\end{eqnarray}
and ${\tilde \mu}(N,H)={\mu}(N,H)-{\mu}_0$ is the oscillatory part of
the chemical potential.
This is a transcendental equation for $\tilde{\mu}(N,H)$ and
   it is difficult to
give ${\tilde \mu}(N,H)$ by solving Eq. (\ref{mutilde})
analytically
except for simple cases such as the single-band case at $T=0$ that we
will discuss below.
After we obtain $\mu(N,H)$, 
we can calculate $M_N(N,H)$ (Eq. (\ref{MNH})).
From the second equation of Eq. (\ref{MNH}) 
and Eq. (\ref{omega0tilde}), 
\begin{eqnarray}
M_{N}(N,H)=-\left( 
\frac{\partial \Omega_0 ( \mu, H)}{\partial H}+
\frac{\partial {\tilde \Omega} ( \mu, H)}{\partial H}
       \right)_{\mu=\mu(N,H)}.
       \label{MN2}
\end{eqnarray}
It is found from Eq. (\ref{domega0/dh}) that
$\partial \Omega_0(\mu,H)/\partial H$
 is independent of
$\mu$. Thus, the first term of of the right side of 
Eq. (\ref{MN2}) 
does not contribute to the oscillation of ${M}_{N}(N,H)$ 
as a function of $1/H$. 
As a result,
we get
\begin{eqnarray}
    \tilde{M}_{N}(N,H)& =& -\left(
     \frac{\partial \tilde{\Omega}(\mu,H)}{\partial H}\right)_{\mu=\mu(N,H)}
\nonumber \\
&=&\tilde{M}_{\mu}(\mu(N,H),H), 
\label{mtilde}
\end{eqnarray}
where $\tilde{M}_{N}(N,H)$ is 
the oscillatory part of $M_{N}(N,H)$.

Recently,
Alexandrov and Bratkovsky\cite{alex3}
asserted that the oscillations of the magnetization also come from
$\Omega_0$ through the oscillation of $\mu$.
Their result for the free
energy (Eq. (\ref{Poisson})
in their paper\cite{alex3}) is formally correct but
they did not take account of the magnetic field dependence
of the $f_\alpha$, which cannot be neglected in two-dimensional systems.
As a result their analysis of the dHvA oscillation for fixed $N$
system (canonical ensemble) is insufficient and their conclusions on
Fourier transform intensities are incorrect.

\begin{figure}[bth]
    \begin{center}
    \leavevmode \epsfxsize=8cm  \epsfbox{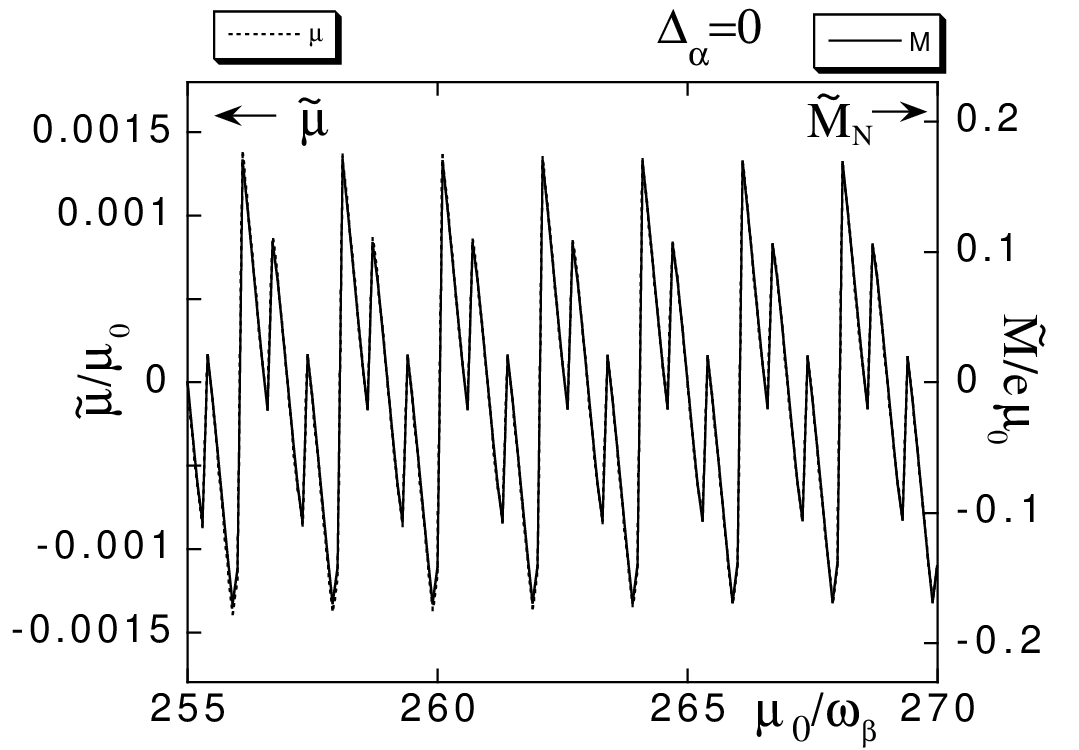}
    \end{center}
   \caption{${\tilde \mu}(N,H)$ (dotted line) and ${\tilde
M}_{N}(N,H)$ (solid line) as a function of $\mu_0/\omega_{\beta}$
when $\Delta_{\alpha}=\Delta_{\beta}=0$.
}
\label{d0}
\end{figure}

\begin{figure}[bth]
    \begin{center}
    \leavevmode \epsfxsize=8cm  \epsfbox{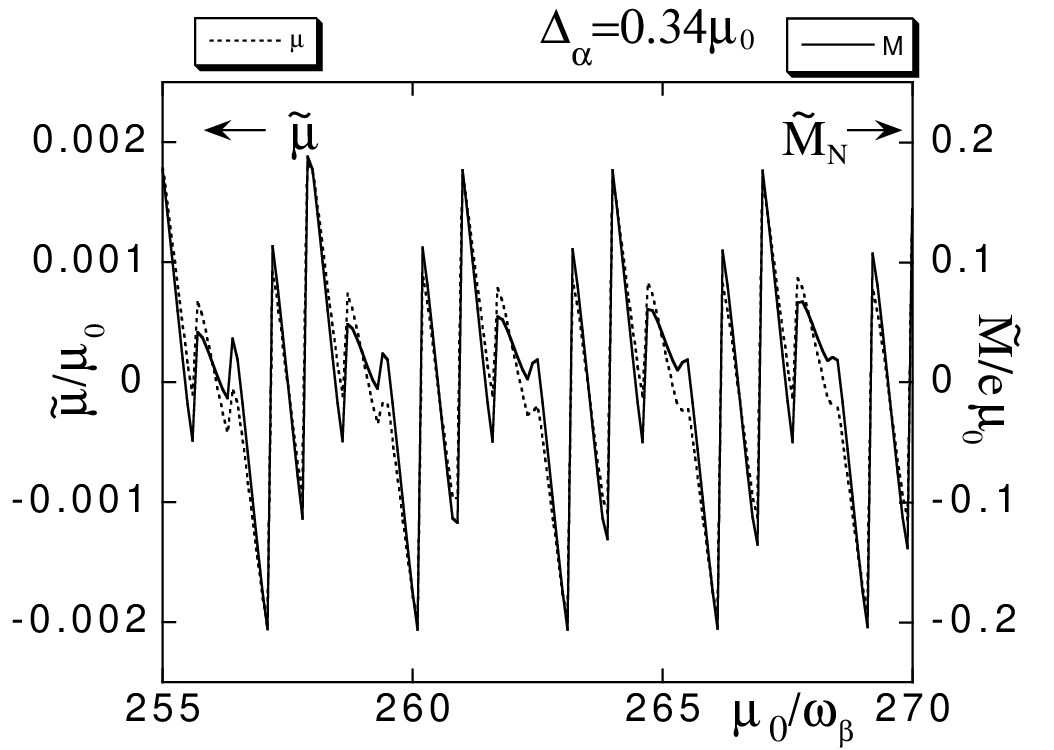}
    \end{center}
   \caption{
The same as Fig.\ref{d0} when $\Delta_\alpha=0.34 \mu_0$
and $\Delta_\beta=0$.
}
\label{d34}
\end{figure}

\begin{figure}[bth]
    \begin{center}
    \leavevmode \epsfxsize=8cm  \epsfbox{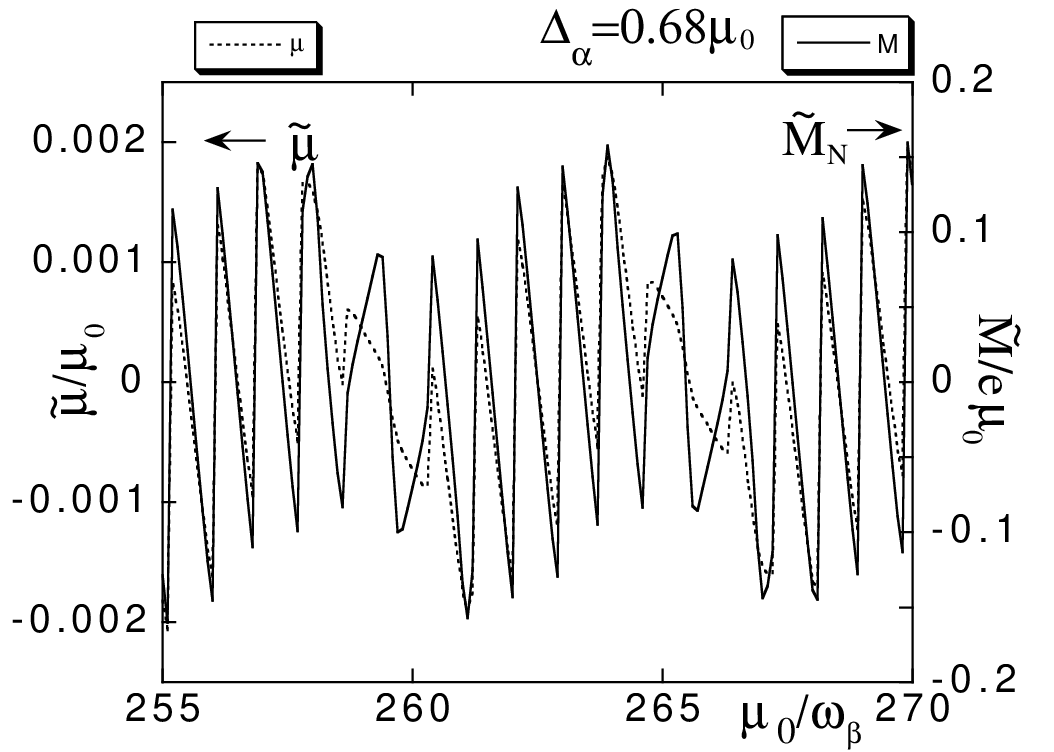}
    \end{center}
   \caption{
The same as Fig.\ref{d0} when $\Delta_\alpha=0.68 \mu_0$
and $\Delta_\beta=0$.
}
\label{d68}
\end{figure}

When all of the band-bottom energies are the same
($\Delta_j=\Delta$), we can derive the simple relation between
$\tilde{\mu}(N,H)$ and $\tilde{M}(N,H)$.
Since $\tilde{\Omega}(\mu,H)$
rapidly oscillates as a function of
$(\mu- \Delta)/H$ in this case
and the envelope depends on $H$ slowly,
we can approximate
\begin{equation}
   \frac{\partial \tilde{\Omega}(\mu, H)}{\partial H}\approx
   - \frac{\partial \tilde{\Omega}(\mu, H)}{\partial \mu }
     \frac{{\mu}(N,H)-\Delta}{H}.
\end{equation}
Using Eq. (\ref{eqNconst}) we get
\begin{eqnarray}
- \frac{\partial \tilde{\Omega}(\mu, H)}{\partial \mu }=N+\frac{\partial
\Omega_0}{\partial \mu},
\end{eqnarray}
and using Eqs. (\ref{totalN}) and (\ref{omega0})
we  get
\begin{eqnarray}
- \frac{\partial \tilde{\Omega}(\mu, H)}{\partial \mu }
  =-2\rho{\tilde \mu}(N,H).
\end{eqnarray}
Then Eq. (\ref{mtilde}) is rewritten
as
\begin{eqnarray}
\tilde{M}_{N}(N,H)&=&\frac{2}{H}\rho{\mu}_0{\tilde \mu}(N,H)
\left(1-\frac{\Delta}{{\mu}_0}+\frac{{\tilde \mu}(N,H)}{{\mu}_0}\right)
\nonumber \\
&\approx&\frac{2\rho({\mu}_0-\Delta)}{H}{\tilde \mu}(N,H)
\nonumber \\
& & \mbox{if $\Delta_j = \Delta$ for all $j$},
\label{MNmu}
\end{eqnarray}
where $|\tilde{\mu}(N,H)/\mu_0| \ll 1 $ is used.
Thus the oscillatory part of the magnetization is proportional to the
oscillatory part of the chemical potential, if all the band-bottom
energies are same.
The similar relation between $\tilde{M}_N(N,H)$ and $\tilde{\mu}(N,H)$
in the single-band system with electron (or hole) reservoirs
has been  obtained independently
by Mineev and Champel\cite{mineev,champel},
Itskovsky{\it et al.}\cite{itsprb} and Grigoriev\cite{grig2}.
However, when the band-bottom energies
are not the same, Eq. (\ref{MNmu}) is not satisfied.
In that case, ${\tilde M}_{N}(N,H)$ should be calculated from
Eq. (\ref{mtilde})
with numerically solved ${\tilde \mu}(N,H)$ in Eq. (\ref{mutilde}).
In order to confirm the above discussion,
we show
${\tilde \mu}(N,H)$ and ${\tilde M}_{N}(N,H)$
in the cases of $\Delta_{\alpha}=\Delta_{\beta}$
and $\Delta_{\alpha}\neq \Delta_{\beta}$
in the two-band system ($j=\alpha, \beta$).
We set $m_{\alpha}=0.5m_0$,
$m_{\beta}=m_0$,
$g_{\alpha}=g_{\beta}=0$, $T=0.0001\mu_0$,
$\Delta_{\beta}=0$ and
$\Delta_{\alpha}=0, 0.34\mu_0$ and $0.68\mu_0$.
By assuming the parabolic band, $\rho_{j}=m_{j}/2\pi$.
The sum over $r$ in Eqs. (\ref{mutilde}) and (\ref{mtilde})
is taken up to 30.
We have checked that the results
are the same within the numerical errors
when we take the sum over $r$ up to 50.
It can be clearly seen from Figs. \ref{d0}
   that ${\tilde \mu}(N,H)$ and $\tilde{M}_N(N,H)$
oscillate  as the same function of $H$ when $\Delta_\alpha=\Delta_{\beta}$.
When $\Delta_{\alpha}$ is not equal to $\Delta_{\beta}$,
   ${\tilde \mu}(N,H)$ is different from
${\tilde M}_{N}(N,H)$, as seen in Figs. \ref{d34} and \ref{d68}.


\begin{figure}[bth]
    \begin{center}
    \leavevmode \epsfxsize=9cm  \epsfbox{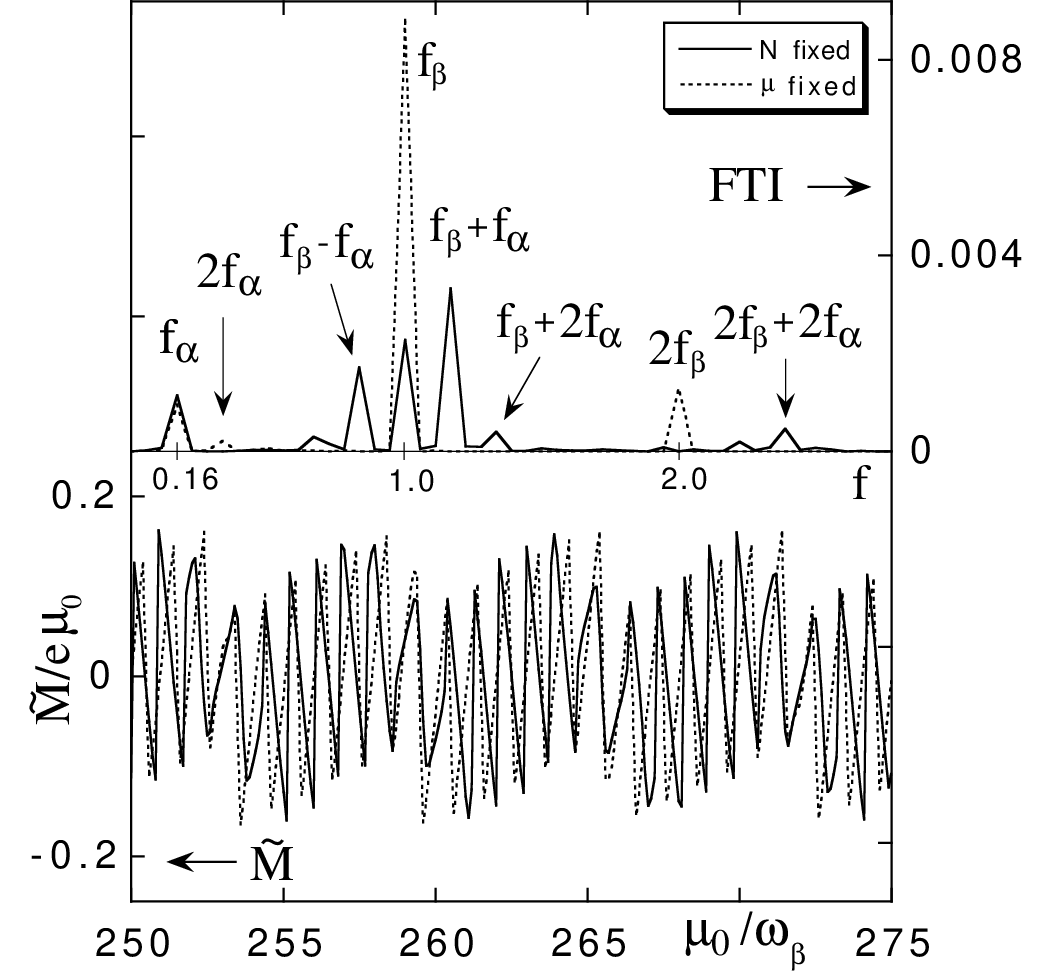}
    \end{center}
   \caption{${\tilde M_{N}(N,H)}$ and
${\tilde M_{\mu}(\mu,H)}$ (lower figures)
as a function of $\mu_0/\omega_{\beta}$ and their Fourier transform
intensities (upper figures), where the frequency has the unit of
$m_0\mu_0/e$. The Fourier transform is
performed in the region of $255\leq \mu_0/\omega_{\beta} \leq 273$.
}
\label{M2band}
\end{figure}

In Fig. \ref{M2band},
we show ${\tilde M}_{N}(N,H)$,
${\tilde M}_{\mu}(\mu,H)$  and their Fourier transform intensities
(FTIs), where we set $m_{\alpha}=0.5m_0$,
$m_{\beta}=m_0$, $\Delta_{\alpha}=0.68\mu_0$,
$\Delta_{\beta}=0$,
$g_{\alpha}=g_{\beta}=0$ and
$T=0.0001\mu_0$.
With these parameters the frequencies are
$f_{\alpha}=m_{\alpha}(\mu_0-\Delta_{\alpha})/e =0.16m_0\mu_0/e$ and
$f_{\beta}=m_0\mu_0/e$.
These parameters are the same as in the previous
studies\cite{nakano,kishigispin2} except for 
the temperature;
   the ground state energy
is calculated directly by filling Landau levels at $T=0$ in the
previous studies.
It is seen from Fig. \ref{M2band} that the
FTI has large peaks at the
combination frequencies $f_{\beta}+f_{\alpha}$,
$f_{\beta}+2f_{\alpha}$, $2f_{\beta}+2f_{\alpha}$ and
$f_{\beta}-f_{\alpha}$
besides at the fundamental frequencies $f_{\beta}$ and $f_{\alpha}$,
which are in agreement with previous
results at $T=0$ \cite{nakano,kishigispin2}.
These combination oscillations are known as the quantum interference
oscillations.
The quantum interference oscillations are caused by the oscillation
of $\mu$.
If $\mu$ is fixed, as seen in dotted line in Fig. \ref{M2band},  the FTI
of $\tilde{M}_{\mu}(\mu,H)$
has peaks at  the
fundamental frequencies ($f_{\beta}$ and $f_{\alpha}$) and
higher harmonics ($2f_{\beta}$ and $2f_{\alpha}$) and it has no peaks at the
combination frequencies.
%

\begin{figure}[bth]
    \begin{center}
    \leavevmode \epsfxsize=8cm  \epsfbox{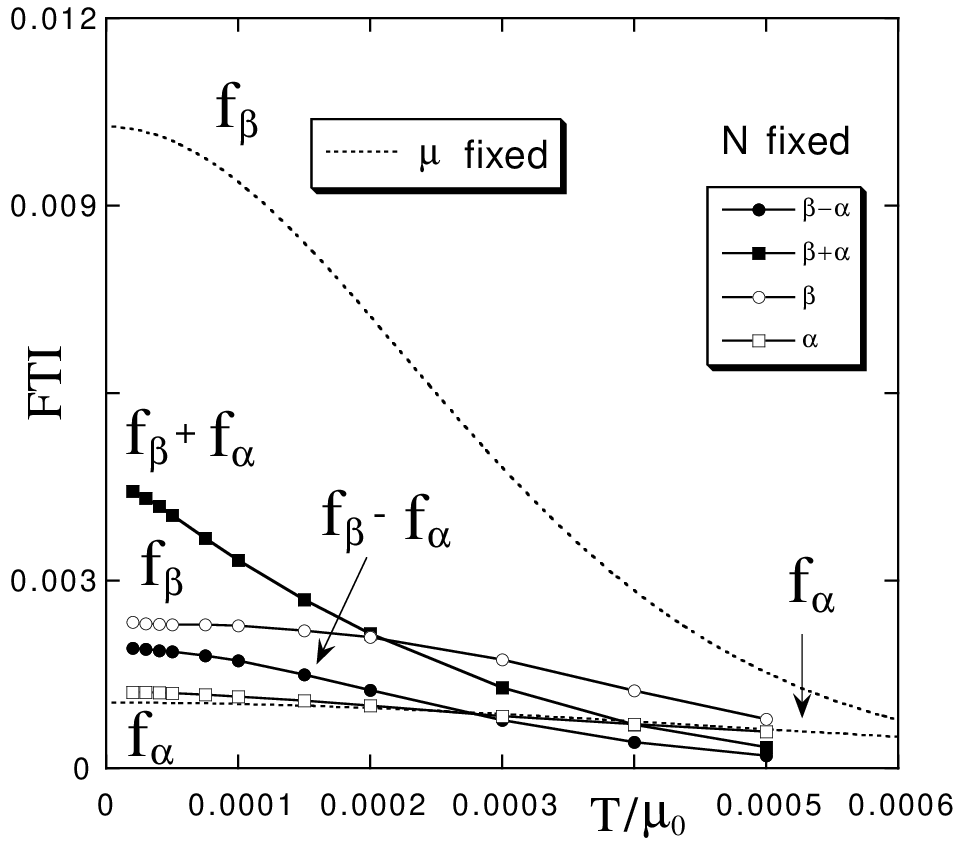}
    \end{center}
   \caption{$T$-dependence of the FTIs of the magnetization
  at $f_{\alpha}$, $f_{\beta}$,
$f_{\beta}-f_{\alpha}$ and $f_{\beta}+f_{\alpha}$
in the fixed $N$ case (solid lines) and FTIs
  at $f_{\alpha}$ and $f_{\beta}$ in the fixed $\mu$ case (dotted lines).
}
\label{temp}
\end{figure}

We show the $T$-dependence of 
the FTIs of $\tilde{M}_{\mu}(\mu,H)$
and $\tilde{M}_{N}(N,H)$
in Fig. \ref{temp}, where the parameters are
the same as those in Fig. \ref{M2band}. 
In order to take in the sharpness of the oscillation at the 
very low temperature, we set the upper limit of 
$r$ in Eqs. (\ref{mutilde}) and (\ref{mtilde}) as 75 below $T/\mu_0 =0.00003$. 
The $T$-dependence of the 
FTI is given by  $R^2_{T,jr}$ 
(Eq. (\ref{RT})) when 
  $\mu$ is fixed.  When $N$ is fixed, the temperature dependence of FTIs
are no longer given by  $R^2_{T,jr}$. Although the temperature
dependence of the FTI at $f_\alpha$ is
well fitted by  $R^2_{T,jr}$, the temperature dependences of the FTIs
at $f_\beta$, $f_\beta \pm f_\alpha$ are not fitted by the temperature
reduction factors,  $R^2_{T,jr}$. Therefore, we should be careful to
determine the cyclotron mass from the 
temperature dependence of the
FTIs in two-dimensional systems.


\subsection{special case: single-band system at $T=0$}
In the single-band ($\rho_j=\rho$, $\omega_j=\omega$,
    $m_j=m$ and $\Delta_j=\Delta$) and no-spin-splitting case
($g_j=0$) at $T=0$,
we can obtain $\tilde{M}_{N}(N,H)$ analytically.
By using the identity
\begin{equation}
    [x] = x-\frac{1}{2} + \sum_{r=1}^{\infty} \frac{1}{\pi r} \sin\left
( 2 \pi r x \right),
\label{eqidentity}
\end{equation}
where $[x]$ is the largest integer satisfying $[x] \leq x$,
Eq.~(\ref{eqNmuh}) at $T=0$ is written as
\begin{equation}
    N(\mu,H) = 2 \rho \omega
     \left[ \frac{\mu-\Delta}{\omega}+\frac{1}{2} \right],
\end{equation}
When we take this equation as a equation giving $\mu(N,H)$, we get
\begin{equation}
    \mu(N,H) -\Delta =\left( \left[ \frac{N}{2 \rho \omega} \right] +
\frac{1}{2} \right) \omega.
\label{eqmuN}
\end{equation}
This equation means that $\mu$ is pinned at the Landau level and it
jumps from one Landau level to the other when
$N/(2 \rho \omega)$ is
integer.
Since $(\mu(N,H)-\Delta)/\omega + 1/2$ is integer,
we obtain from Eq. (\ref{eqtildeomega})
that
$\tilde{\Omega}(\mu(N,H),H)$ does not oscillate as a function of $H$
in the single-band case at $T=0$
when $N$ is fixed.
By using the identity (Eq.~(\ref{eqidentity})) in Eq.~(\ref{eqmuN}),
we get
\begin{equation}
{\mu}(N,H)=\frac{N}{2\rho} + \Delta
    + \sum_{r=1}^{\infty} \frac{\omega}{\pi r}\sin \left
( 2 \pi r \frac{f}{H} \right),
\label{mu_single}
\end{equation}
where $ f = N m/(2\rho e)$.

It should be noticed that 
if Eq.~(\ref{eqNmuh}) is considered as an 
equation giving $\mu(N,H)$ in the fixed $N$ case, 
this equation is not satisfied when 
Eq.~(\ref{eqmuN}) is inserted: 
If we insert 
Eq.~(\ref{eqmuN}) in Eq.~(\ref{eqNmuh}), we
 get the right hand side as $2 \rho (\mu(N,H) - \Delta)$ 
for the single-band case, because 
$(\mu(N,H) - \Delta)/\omega 
+1/2$ is integer in Eq.~(\ref{eqmuN}). 
Although $\mu(N,H)$ oscillates as a function of 
$H$, the left hand side of Eq.~(\ref{eqNmuh}) 
should be independent of $H$ when $N$ is 
fixed. 
Therefore, we cannot insert Eq.~(\ref{eqmuN})
 obtained at $T=0$ into Eq.~(\ref{eqNmuh}). 
The reason for this paradoxical result is the following. 
The Landau level 
with the energy $\mu(N,H)$ is partially filled at $T=0$. 
This information is lost in Eq.~(\ref{eqmuN}). 
In the same way we cannot get $\tilde{M}_{N}(N,H)$
if we first set
$T=0$ to get $\mu(N,H)$ and insert it to get
$\tilde{M}_{N}(N,H)=\tilde{M}_{\mu}(\mu(N,H),H)$ in
Eq. (\ref{eqmtildemu}).

The relation Eq. (\ref{MNmu})
is always satisfied in the single-band case even 
at $T=0$. 
From Eqs. (\ref{MNmu}) and (\ref{mu_single}), 
we get
\begin{equation}
     \tilde{M}_{N}(N,H) = \frac{eN}{m}
\sum_{r=1}^{\infty} \frac{1}{\pi r}\sin \left
( 2 \pi r \frac{f}{H} \right).
\label{eqmtildeN}
\end{equation}
We see from Eq. (\ref{eqmtildemu}) and Eq. (\ref{eqmtildeN}) that the
oscillation of the magnetization for the fixed $N$ has a phase-shifted and
sign-changed saw-tooth shape comparing with that obtained
in the fixed
$\mu$ in the two-dimensional single-band spinless case at $T=0$.


\section{quasi-two-dimensional tight-binding electrons}\label{sect:q2d}

In this section, we
study the crossover from two-dimension to three-dimension in
   the quasi-two-dimensional electrons at
$T=0$.
If we do not take into
account the periodicity of the system, the system has the closed Fermi surface
even for the very large effective mass in the $z$-direction.
Therefore, in order to study the crossover from two-dimension
(with  cylindrical Fermi surface) to
three-dimension we have to study the periodic system.

\begin{figure}[bth]
    \begin{center}
    \leavevmode \epsfxsize=8cm  \epsfbox{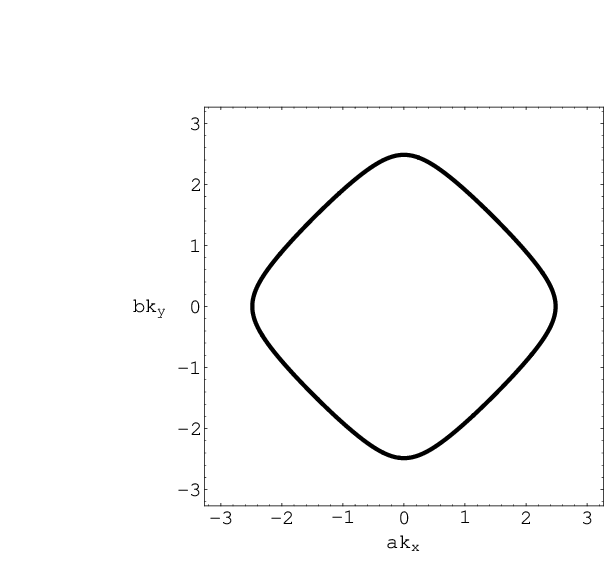}
    \end{center}
   \caption{Fermi surface of the two-dimensional tight-binding electrons
with filling factor 1/3.
}
\label{figfs1}
\end{figure}

We take the tight-binding electrons in the simple orthogonal
lattice with lattice constants $a$, $b$ and $c$ and the hoppings
$t_x$, $t_y$ and $t_z$. For simplicity we neglect the spin
(i.e. $g=0$) and
set $t_x=t_y=t$.
The Hamiltonian in the absence of the magnetic field is given
by
\begin{equation}
   \hat{\cal H}_{0} =
   \sum_{\bf k} \hat{c}^{\dagger}({\bf k}) {\cal E}({\bf k})
\hat{c}({\bf k}),
\label{hamiltonian0}
\end{equation}
where
\begin{equation}
   {\cal E}({\bf k})=- 2t (\cos a k_x + \cos b k_y) -  2t_z \cos c  k_z.
\end{equation}
The magnetic field is introduced by the Peieres substitution,
\begin{equation}
   {\bf k} \rightarrow {\bf k} + e {\bf A},
\end{equation}
where ${\bf A}$ is a vector potential.
In this paper the magnetic field is taken to be along the $z$-direction and
the Landau gauge is used,
\begin{equation}
   {\bf A}= (H y, 0,0) .
\end{equation}
Then the Hamiltonian is written as
\begin{eqnarray}
\hat{\cal H} &=&  -t \sum_{{\bf k}} \left\{
               \exp(i a k_{x})\hat{c}^{\dagger}
    (k_{x},k_{y}-\delta,k_z)\hat{c}({\bf k})
                       +{\rm h.c.} \right\}
                          \nonumber  \\
&+& \sum_{{\bf k}} \left( -2t\cos( b k_{y}) -2t_{z}
    \cos k_{z}\right)\hat{c}^{\dagger}({\bf k})\hat{c}({\bf k}).
\label{e:om}
\end{eqnarray}
where
\begin{equation}
   \delta =\frac{eaH}
{\hbar c_{0}}=\frac{\phi}{\phi_0}\frac{2\pi}{b},
\end{equation}
$\phi=abH$ is the flux passing through a unit cell, and
$\phi_{0}=2\pi\hbar c_{0}/e$ is the unit flux quantum.
For the numerical calculation we take
the number of the flux quantum per plaquette,
\begin{equation}
   h = \frac{\phi}{\phi_0}
\end{equation}
to be a rational number. We use $h$ instead of $H$ hereafter.
If we take  $a=b\approx
10$ [\AA] (the  typical values for quasi-two-dimensional organic
conductors),  $h\approx 1/40$ corresponds to $H=100$ [T].
We obtain the eigenvalues ($\epsilon_{i}(h)$)
by diagonalizing $\hat{\cal H}$ numerically.
Then we calculate the Helmholtz free energy $F(N,h)$ at $T=0$,
which is just the total energy, as
\begin{equation}
F(N,h) = \frac{1}{N_s}\sum_{i=1}^{N}
\epsilon_{i}(h),
\end{equation}
where $N_s$ is the site number.
We fix the electron number to be $1/3$ filled, i.e. $N/N_s=1/3$.
The Fermi surface at $h=0$ and $t_z=0$ is shown in Fig.~\ref{figfs1}.
The area of the closed Fermi surface is
\begin{eqnarray}
   f = \frac{1}{3}f_{\rm BZ},
\nonumber
\end{eqnarray}
where $f_{\rm BZ}=4\pi^2/ab$.
The magnetization is obtained from the numerical differential
\begin{equation}
   M_N(N,h) = -\frac{\delta F(N,h)}{\delta h}.
\end{equation}
The chemical potential is given by
\begin{equation}
\mu(N,h)=\epsilon_{N}(h).
\end{equation}

\begin{figure}[tb]
    \begin{center}
    \leavevmode \epsfxsize=8cm  \epsfbox{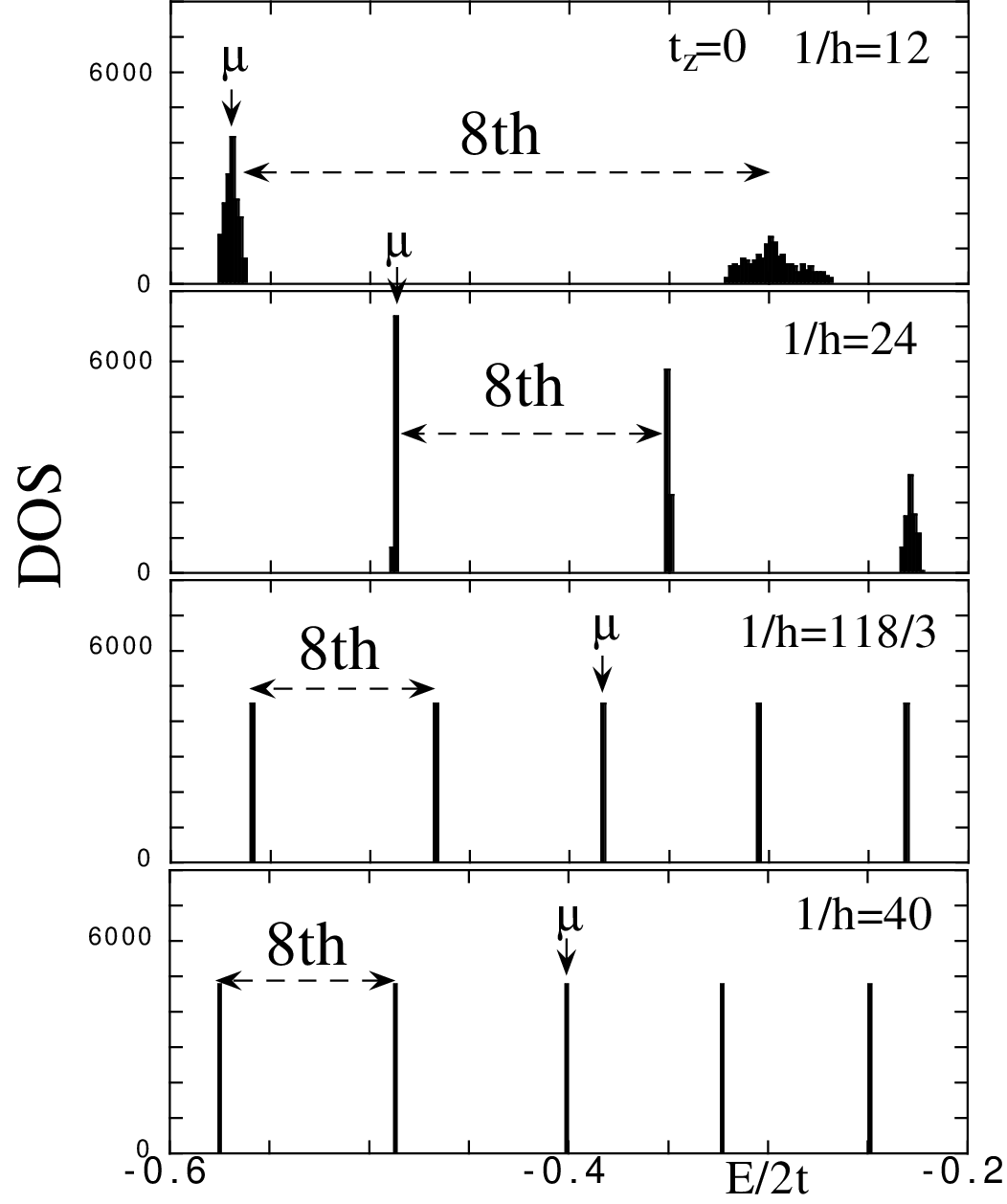}
    \end{center}
   \caption{Density of states for the
tight-binding electrons in the magnetic field
in the two-dimensional system ($t_z=0$).  The arrows show the
chemical potential for 1/3-filled band.
}
\label{figdos}
\end{figure}

In the tight-binding model, even in the two-dimensional case ($t_z=0$),
the density of states is not the sum of the delta functions but it consists
of mini-bands with the finite width (Harper broadening). However, the width
is very small especially in the case of $t_a=t_b$\cite{hasegawa90} as shown
in Fig.~\ref{figdos}.
On the other hand, the mini-gap, which means the spacing between
the Landau levels,
is of the order of $8th$, as indicated by dotted lines in
Fig.~\ref{figdos}.
(The total band width is of the order of $8t$,
which consists of $q$
delta-function-like mini-bands
when $h=p/q$ with co-prime integers $p$ and $q$.
However, $p$ mini-bands are very close each other, making one super-mini-band.
Therefore, the mini-gap is roughly of the
order of $8t/(q/p)=8th$.
For example, consider
$h=1/40$ and $h=3/118$. These two cases should have similar density of
states as can be seen in Fig.~\ref{figdos}. 
Indeed, they have the similar mini-band-gap,
$8t \times (1/40)\approx 8t \times (3/118)$,
if we neglect the mini-mini-gap inside the super-mini-band which
consists of three mini-bands in the case of $h=3/118$.)

\begin{figure}[tb]
    \begin{center}
    \leavevmode \epsfxsize=8cm  \epsfbox{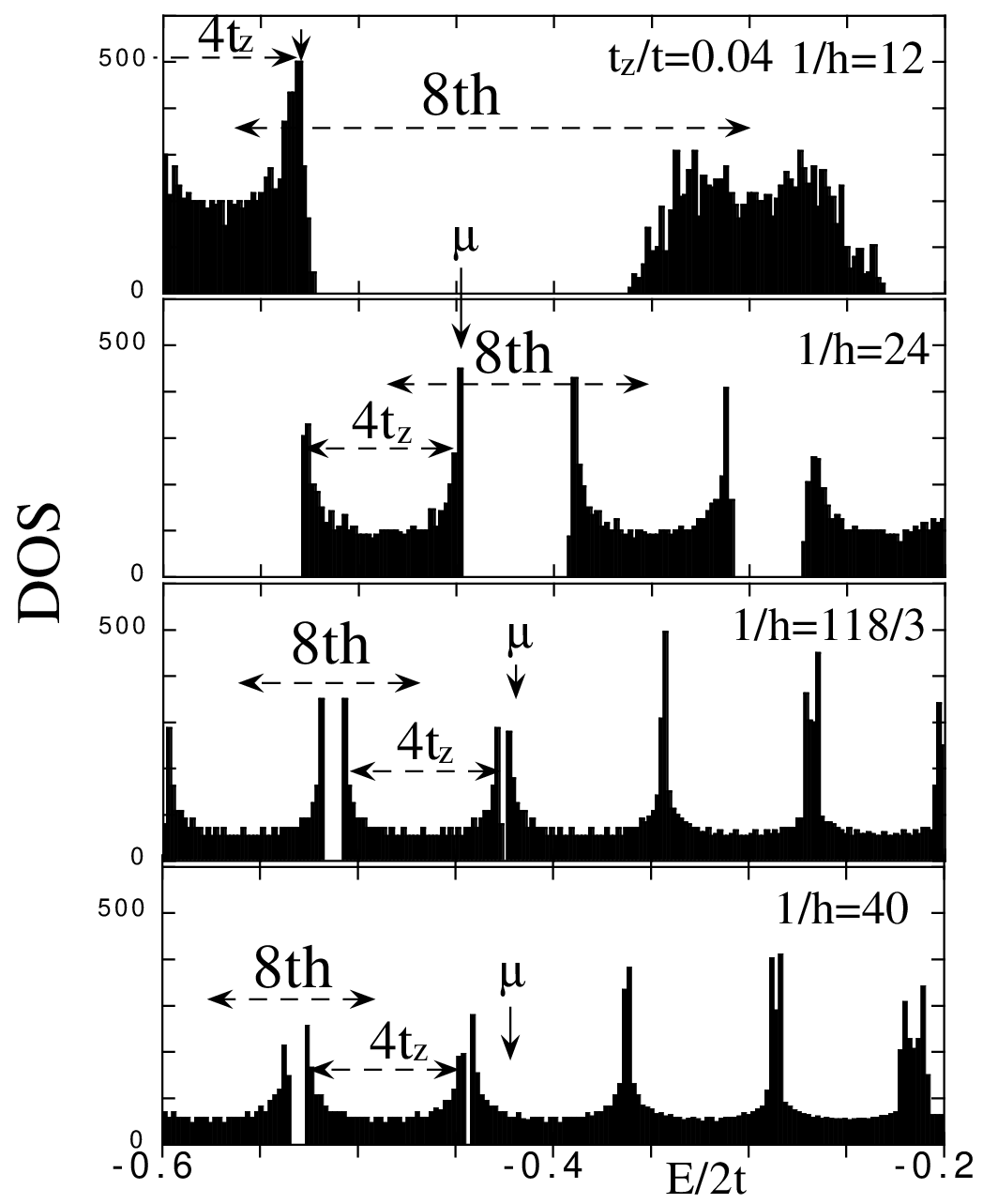}
    \end{center}
   \caption{Density of states for the
tight-binding electrons in the magnetic field
in the quasi-two-dimensional system
($t_z/t=0.04$).  The arrows show the
chemical potential for 1/3-filled band.
}
\label{figdosz}
\end{figure}

When $t_z$ becomes finite, the density of states
becomes broader as shown in Fig.~\ref{figdosz}.
The width of each mini-band becomes the order of $4t_z$.
If the width of the mini-band is the same order of the mini-gap
at $t_z=0$, that is, $4t_z \approx 8th$,
the density of states has no energy-gap as in the three-dimensional systems
in the magnetic field.
We expect the crossover from two-dimension to
three-dimension occurs at $h \approx t_z/2t$.
In fact, we can see this crossover from the density of states
for $t_z/t=0.04$ and $1/h=40$, as shown
in Fig.~\ref{figdosz}.
The actual dimensional crossover in the quasi-two-dimensional organic
conductors may occur at $H \approx 20$ [T].
As, in these organic conductors, $8t$ is of the order of 1
[eV] and $t_z$ is of the order of 10 [K]\cite{hanasaki,uji2},
the crossover field is $h \approx t_z/2t\approx 1/200$,
that is, $H \approx 20$ [T].

\begin{figure}[tb]
    \begin{center}
    \leavevmode \epsfxsize=8.5cm  \epsfbox{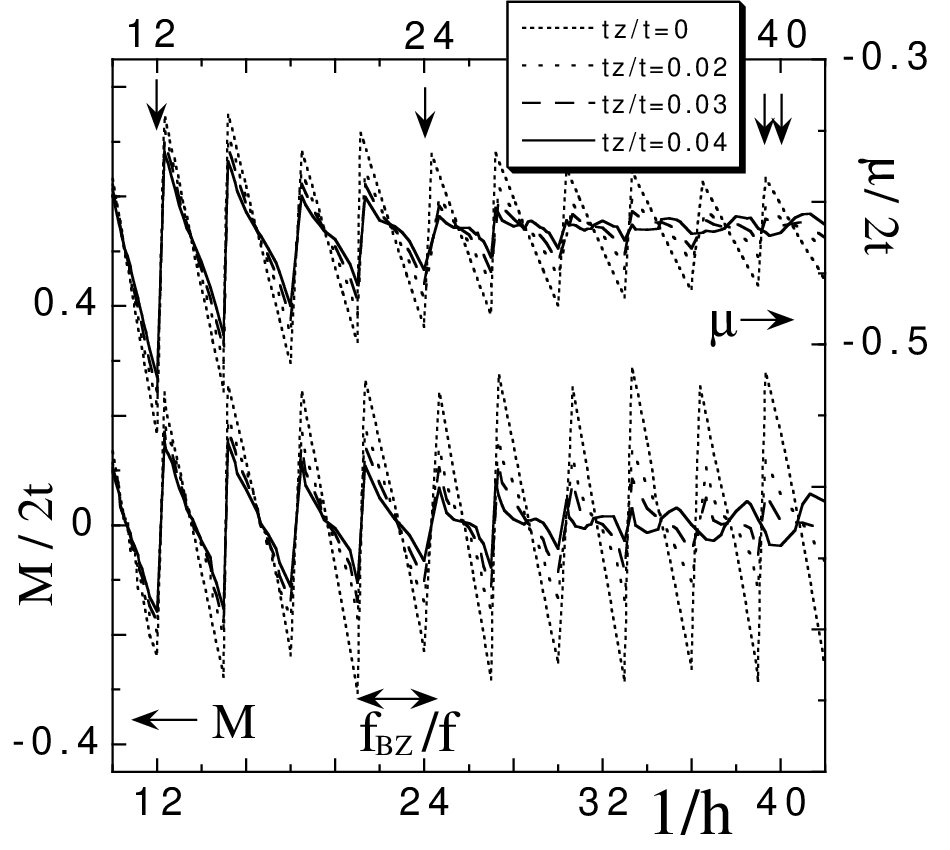}
    \end{center}
   \caption{$\mu$ (upper figure) and $M_N(N,h)$
(lower figure) as a function of $1/h$.
Arrows indicate the magnetic field for which the density of states are
plotted in Figs.~ \ref{figdos} and \ref{figdosz}.
}
\label{figmu}
\end{figure}

When the system is two-dimensional ($t_z=0$),
the chemical potential is pinned in the narrow region of
the non-zero density of states as in the case
of the two-dimensional free electrons.
In this case, the chemical potential and the magnetization
oscillate periodically as a function of $1/h$,
as shown by the dotted lines in Fig.~\ref{figmu}.
The frequency of the oscillation of
$\mu(N,h)$ and $M_N(N,h)$ is $1/3$, which is in agreement
with the closed Fermi surface area in
the unit of the Brillouin zone, $f/f_{\rm BZ}$.

In the quasi-two-dimensional system, the amplitudes of the
oscillations of $\mu(N,h)$ and $M_N(N,h)$ become smaller as
$t_z$ increases as seen in Fig.~\ref{figmu}.
In particular, the oscillation of $\mu(N,h)$ is very small
when the density of  states becomes gapless ( $h=1/40$
and $t_z/t=0.04$ as shown in Fig.~\ref{figdosz},
  where $4t_z$ is nearly equal  to $8th$, i.e.,
$4t_z/(8t h)=t_z/(2th) =0.8$).

%

\begin{figure}[t]
    \begin{center}
    \leavevmode \epsfxsize=8.5cm  \epsfbox{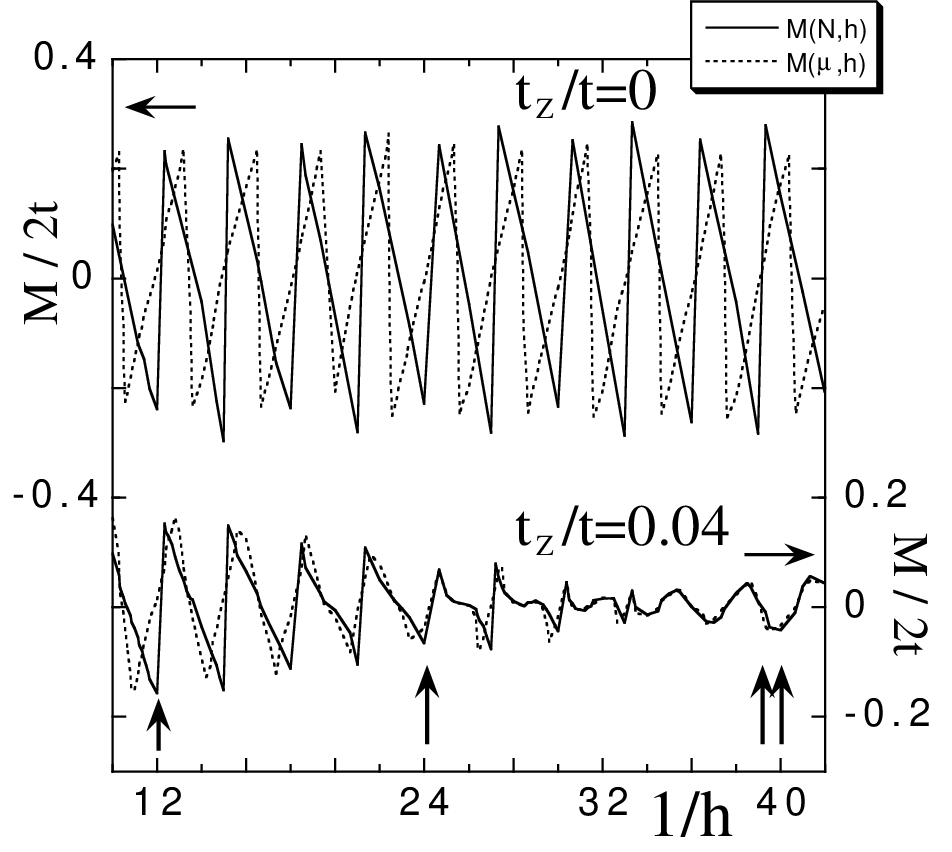}
    \end{center}
   \caption{
$M_ N(N,h)$ and $M_ {\mu}(\mu,h)$ as a function of $1/h$.
Arrows indicate the magnetic field for which the density of states are
plotted in Figs.~ \ref{figdos} and \ref{figdosz}.
}
\label{figM}
\end{figure}

We compare the magnetizations in the fixed $N$ and fixed $\mu$ cases in
Fig.~\ref{figM}.
   The magnetization with fixed $\mu$ at $T=0$ is obtained from the
numerical differential of the thermodynamic
potential $\Omega(\mu,h)$ as
\begin{equation}
    M_{\mu}(\mu,h) = -\frac{\delta \Omega(\mu,h)}{\delta h} ,
\end{equation}
where
\begin{eqnarray}
\Omega(\mu,h)&=&\frac{1}{N_s}\sum_{\epsilon_{i}\leq \mu}
\left( \epsilon_{i}(h) -\mu \right).
\end{eqnarray}
The chemical potential is fixed to be $\mu/2t=-0.414$ to satisfy
$N/N_s=1/3$ at $h=0$.
In Fig.~\ref{figM} the magnetization is plotted as a function of $1/h$
for both cases of  fixed $\mu$ and  fixed $N$.
We calculate $M_N(N,h)$ and $M_{\mu}(\mu,h)$ for two-dimensional case
($t_z=0$) and quasi-two-dimensional case ($t_z/t=0.04$).
As expected, for the
two-dimensional case ($t_z=0$), two curves have inverted and shifted
saw-tooth shape each other, which can be seen from
the upper figures in Fig.~\ref{figM}.

In the quasi-two-dimensional case, the difference between $M_N(N,h)$ and
$M_{\mu}(\mu,h)$ can be seen at $h \gtrsim 1/24$ and it becomes small
at smaller field as clearly seen in Fig.~\ref{figM}. This is consistent with
the behavior of $\mu(N,h)$ shown in
Fig.\ref{figmu}, i.e. the amplitude of the oscillation of $\mu(N,h)$ becomes
small when
$t_z/(2t) \gtrsim h$.

\section{Quantum interference oscillation in quasi-two-dimensional
electrons}\label{QIO}
\begin{figure}[bt]
    \begin{center}
    \leavevmode \epsfxsize=8cm  \epsfbox{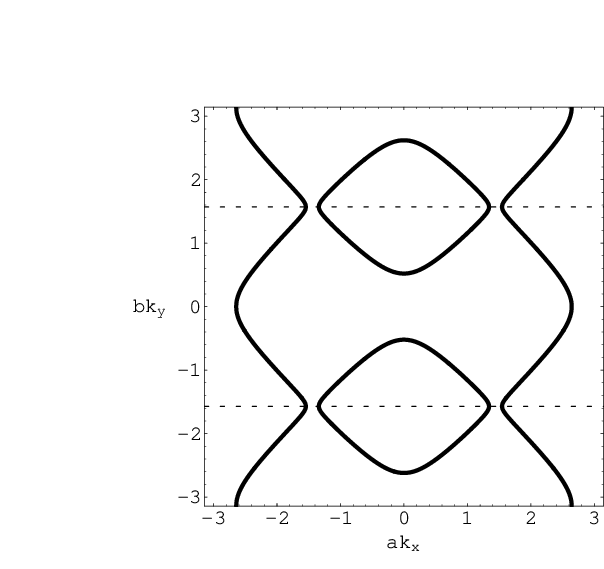}
    \end{center}
   \caption{Fermi surface of the two-dimensional tight-binding electrons
in the presence of periodic potential
with filling factor 7/18. The area of the electron pocket centered at
$(0, \pm \pi/2)$ is $f_\alpha /f_{\rm BZ}\approx 1/18$ in the unit of the 
area of
the extended Brillouin zone ($(2\pi)^2$). The orbit enabled by the
magnetic breakdown has the area $f_\beta /f_{\rm BZ} \approx 7/18$.
}
\label{figfermimb}
\end{figure}

In this section we study the dHvA oscillation in the quasi-two-dimensional
systems
which has the small cylindrical Fermi surface and the quasi-one-dimensional
Fermi surface, as shown in
Fig.~\ref{figfermimb}.
This Fermi surface is realized in $\kappa$-(BEDT-TTF)$_2$Cu(NCS)$_2$,
$\alpha$-(BEDT-TTF)$_2$KHg(SCN)$_4$ and GaAs/AlAs heterointerface.
The small cylindrical Fermi surface is separated from the
quasi-one-dimensional Fermi surface
on the first Brillouin zone edge
(dotted lines in Fig. \ref{figfermimb})
by the periodic potential.
In the semiclassical picture, electrons are expected
to execute a large closed orbital motion
by tunneling the first Brillouin
zone gap at high fields.
Then, the oscillation due to the large closed orbit appears in the dHvA
oscillation.
This is known as the magnetic breakdown phenomena. In
$\kappa$-(BEDT-TTF)$_2$Cu(NCS)$_2$
and $\alpha$-(BEDT-TTF)$_2$KHg(SCN)$_4$, the magnetic breakdown is observed.

%
In order to obtain this Fermi surface we add the periodic potential
\begin{eqnarray}
\hat{\cal V} &=& V \sum_{{\bf r}=(a i_x, b i_y, c i_z)}
(-1)^{i_y} \hat{c}^{\dagger}({\bf r}) \hat{c}({\bf r}) \nonumber \\
   &=& V \sum_{\bf k} \left\{
   \hat{c}^{\dagger}({\bf k}+{\bf g}) \hat{c}({\bf k}) + {\rm h.c.}
\right\},
\label{hamV}
\end{eqnarray}
where
\begin{equation}
   {\bf g}= (0, \frac{\pi}{b}, 0),
\end{equation}
to the Hamiltonian Eq. (\ref{hamiltonian0}) (when $h=0$) or
Eq. (\ref{e:om}) (when $h \neq 0$).
When $V \not= 0$ and $h=0$, the Brillouin zone is folded, i.e.
$k_{x}\in[-\pi/a, \pi/a]$, $k_{y}\in[-\pi/2b, \pi/2b]$ and
$k_{z}\in[-\pi/c, \pi/c]$.
In this section, we set electron filling factor $N/N_s$ to be $7/18$ and
$V/2t=0.1$.
In this case, $\mu/2t \approx -0.254$ when $h=0$.
In Fig.~\ref{figfermimb} we plot the Fermi surface at $t_{z}=0$ and $h=0$.
The small electron pocket has the area
\begin{eqnarray}
   f_{\alpha} \approx \frac{1}{18}f_{\rm BZ}
\nonumber
\end{eqnarray}
and the large orbit enabled by the magnetic breakdown at the gap on
the zone edge has the area
\begin{eqnarray}
   f_{\beta} \approx \frac{7}{18}f_{\rm BZ}.
\nonumber
\end{eqnarray}

\begin{figure}[t]
    \begin{center}
    \leavevmode \epsfxsize=9cm  \epsfbox{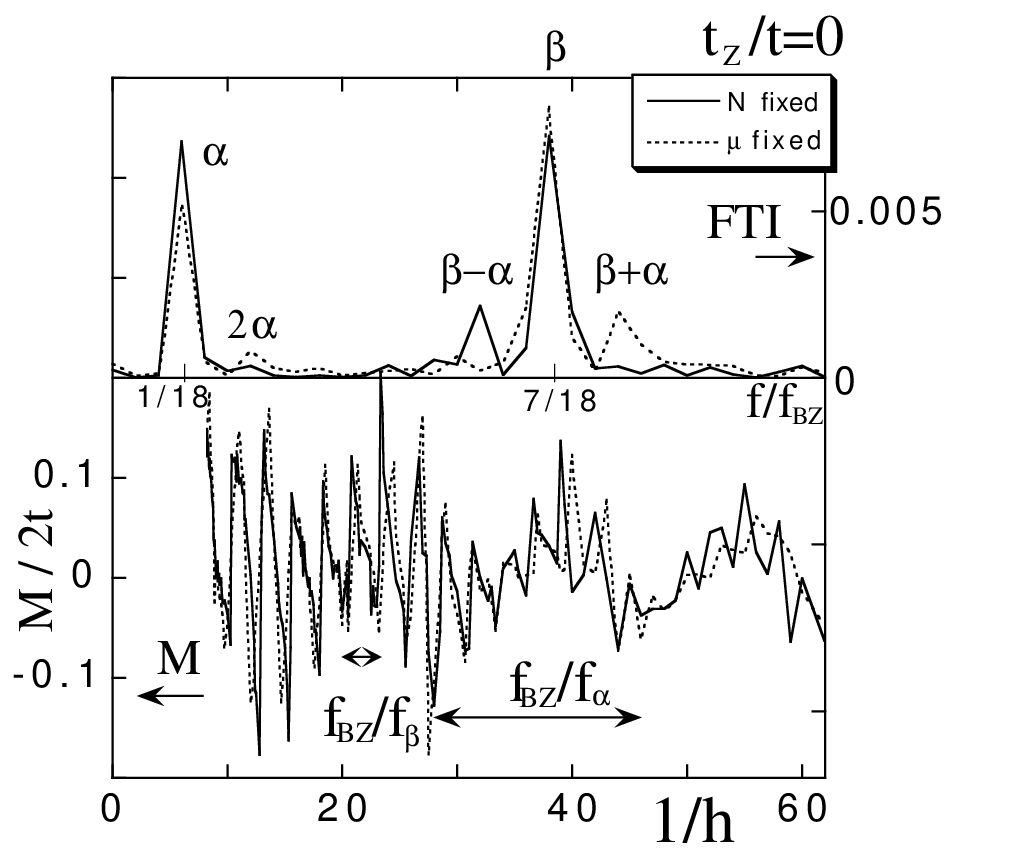}
    \end{center}
   \caption{$M_N(N,h)$ and $M_{\mu}(\mu,h)$ (lower figure) as a function of
$1/h$ and their
Fourier transform intensities (upper figure) at $t_z=0$. The Fourier
transform is
performed in the region of  $12\leq 1/h \leq 62$.
The peaks of the FTI occur at $f_{\alpha}$, $f_{\beta}$,
$f_{\beta}-f_{\alpha}$ and $f_{\beta}+f_{\alpha}$ besides smaller peaks at
$2f_\alpha$, $f_{\beta} \pm 2 f_{\alpha}$ etc.
}
\label{fig0mb}
\end{figure}

We calculate the magnetization and its FTIs numerically as in the
previous section. In Fig.~\ref{fig0mb} we show the results in the
two-dimensional case.
It can be seen from the lower figure in Fig.~\ref{fig0mb}
    that the amplitude of the oscillation with the frequency $f_\beta$
    becomes large due to magnetic breakdown as $1/h$ becomes small.
In the magnetic breakdown system, the oscillation with
$f_\beta + f_\alpha$
is expected to exist but the oscillation with
$f_\beta - f_\alpha$ is forbidden in the
semiclassical picture. This semiclassical picture is consistent with our
quantum calculation as seen in the dotted line in the upper figure in
Fig. \ref{fig0mb},
as known previously\cite{harrison,kishigi2}.
If the two bands are not connected with magnetic breakdown,
combination frequencies
($f_\beta \pm f_\alpha$ etc.) do not
appear in $M_{\mu}(\mu,H)$,
as can be seen from dotted lines in Fig. \ref{M2band}.
On the other hand, in the two-band system, both frequencies $f_\beta
\pm f_\alpha$ are seen in $M_N(N,H)$ (Fig. \ref{M2band}), where
the FTI at
$f_{\beta} + f_\alpha$ is larger than that at $f_{\beta} - f_\alpha$.
In the magnetic breakdown system, however, the FTI at
$f_\beta+f_\alpha$ in $M_N(N,H)$
is very small as seen in the solid line
in the upper
figure in Fig.~\ref{fig0mb}.
This behavior has also been known previously\cite{harrison,kishigi2}.

In the previous papers we have shown that the FTI at
$f_\beta + f_\alpha$ in $M_N(N,H)$
in the magnetic breakdown system is
anomalously {\em enhanced} by spin-splitting
effect\cite{kishigispin,kishigispin2}. We have
interpreted the anomalous spin-enhancement of the FTI
at $f_\beta + f_\alpha$ in $M_N(N,H)$
in the magnetic breakdown system as
follows.
    From the results of
two-band and magnetic breakdown systems,
it is found that the oscillation with $f_\beta + f_\alpha$ is caused
by both effects of the
magnetic breakdown
and the chemical potential oscillation, as can be seen in the fixed
$\mu$ case in Fig.~\ref{fig0mb} and in the
fixed $N$ case in Fig. \ref{M2band}, respectively.
If the electron spin is neglected
in  magnetic breakdown systems
in the fixed $N$ case, these two origins of the $f_\beta + f_\alpha$
oscillation  almost cancel each other resulting
in the small FTI at
$f_\beta + f_\alpha$ in $M_N(N,H)$.
The spin-splitting may reduce the chemical potential
oscillation.
As a result, the cancellation becomes imperfect and the FTI at
$f_\beta + f_\alpha$ in $M_N(N,H)$
in the magnetic breakdown system is enhanced due to the spin-splitting.

\begin{figure}[t]
    \begin{center}
    \leavevmode \epsfxsize=9cm  \epsfbox{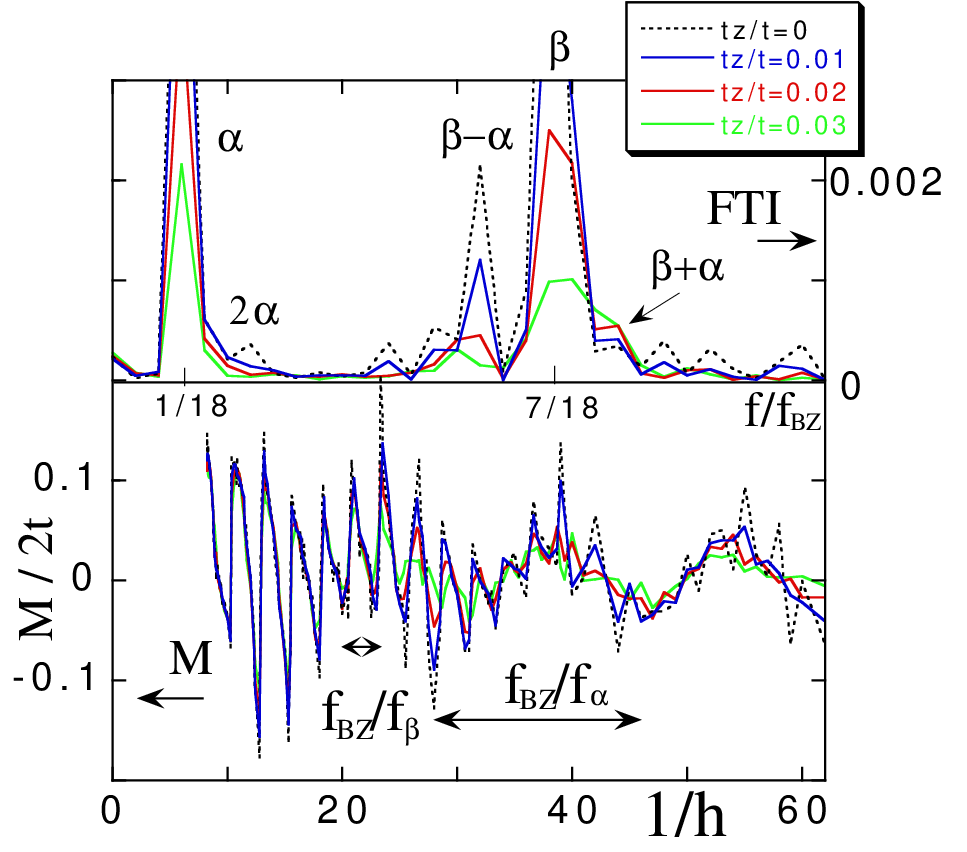}
    \end{center}
   \caption{$M_N(N,h)$ (lower figure) as a function of $1/h$ and their
Fourier transform intensities (upper figure). The Fourier transform is
performed in the region of  $12\leq 1/h \leq 62$.
The peaks of the FTI occur at $f_{\alpha}$, $f_{\beta}$,
$f_{\beta}-f_{\alpha}$ and $f_{\beta}+f_{\alpha}$ besides smaller peaks at
$2f_\alpha$, $f_{\beta} \pm 2 f_{\alpha}$ etc.
}
\label{figmmb}
\end{figure}

\begin{figure}[t]
    \begin{center}
    \leavevmode \epsfxsize=8cm  \epsfbox{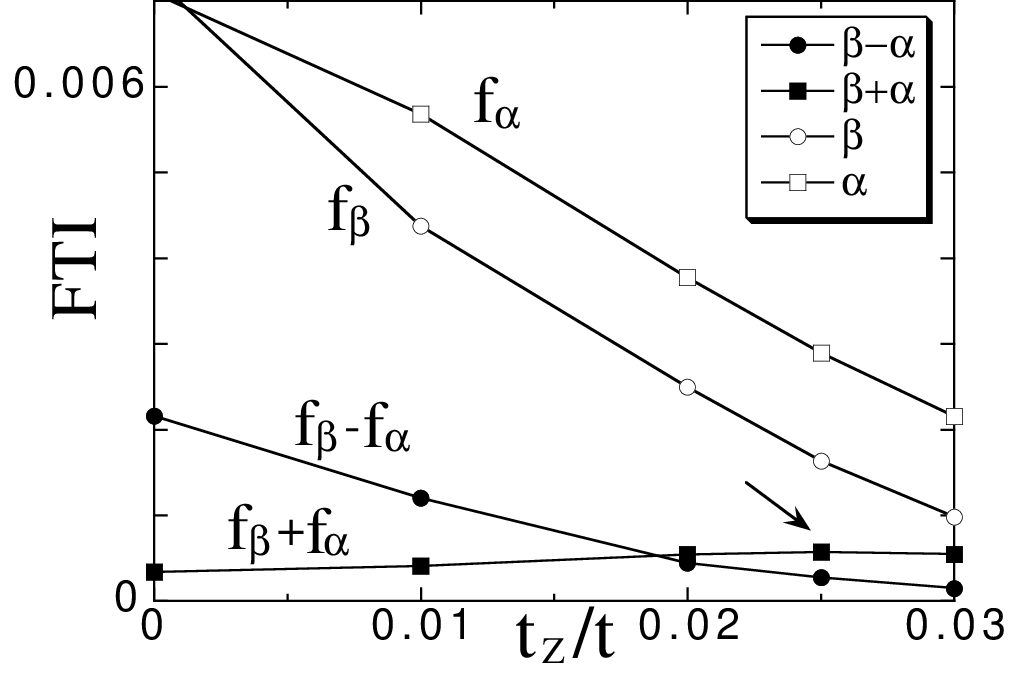}
    \end{center}
   \caption{Main FTIs of $M_N(N,h)$ as a function of $t_z$. An arrow
indicates a maximum.
}
\label{figmtz}
\end{figure}

The similar effect is expected in the case of quasi-two-dimensional system.
As $t_z$ is increased, the chemical potential oscillation is suppressed,
resulting in the {\em enhancement} of the FTI at
$f_\beta + f_\alpha$.
This is indeed the case as we show in Figs. \ref{figmmb} and \ref{figmtz}.
In Fig. \ref{figmmb} we plot
$M_{N}(N,h)$ and its FTIs for
$t_z/t$=0, 0.01, 0.02 and 0.03.
As $t_z$ increases, the FTIs at $f_\alpha$, $f_\beta$ and $f_\beta -
f_\alpha$ decrease as in the single-band case
studied in the previous
section. The FTI at $f_\beta + f_\alpha$, however, depends very weakly on
$t_z$; it gradually increases as $t_z$ increases, and it has a maximum
around $t_z/t \approx 0.025$ as shown by the arrow in Fig. \ref{figmtz}.



\section{Conclusion}
We find that the dHvA oscillation in  two-dimensional
multi-band systems is given by
Eq. (\ref{mtilde}) with the chemical potential
obtained by solving Eq. (\ref{mutilde}).
It is the general formula,
i.e., it can be applied in  the multi-band systems at finite
temperatures including the electron spin.
In
two-dimensional systems,
  the cyclotron effective mass is generally not obtained by fitting the
temperature dependence with the reduction factor $R_{T,jr}$.
It should be obtained from the fitting by
Eq. (\ref{mtilde}) with Eq. (\ref{mutilde}).

We also study the crossover of the dHvA oscillation from two-dimensional
systems to three-dimensional systems.
In the simple case of the single Fermi surface we show that the
oscillation of the magnetization calculated
in the canonical ensemble
(fixed $N$) approaches smoothly to that calculated in the
grand canonical ensemble (fixed $\mu$).
In the magnetic breakdown system
we find that the Fourier transform
intensity at $f_{\beta}+f_{\alpha}$ gradually increases as a
three-dimensionality increases and it has a maximum
at $t_z/t \approx 0.025$.
The $t_z$ dependence can be observed experimentally
by applying the uniaxial stress in
the quasi-two-dimensional organic conductors\cite{uniaxial,uniaxial2},
since the uniaxial stress along the $z$-axis
will increase $t_z$.

\section{Acknowledgments}
One of the authors (K.K.) would like to thank T. Yanagisawa and
K. Yamaji for useful discussions.

\end{document}